# A Review of Situation Awareness Assessment Approaches in Aviation Environments

Thanh Nguyen, Chee Peng Lim, Ngoc Duy Nguyen, Lee Gordon-Brown, Saeid Nahavandi, *Senior Member, IEEE*

*Abstract*—Situation awareness (SA) is an important constituent in human information processing and essential in pilots' decision-making processes. Acquiring and maintaining appropriate levels of SA is critical in aviation environments as it affects all decisions and actions taking place in flights and air traffic control. This paper provides an overview of recent measurement models and approaches to establishing and enhancing SA in aviation environments. Many aspects of SA are examined including the classification of SA techniques into six categories, and different theoretical SA models from individual, to shared or team, and to distributed or system levels. Quantitative and qualitative perspectives pertaining to SA methods and issues of SA for unmanned vehicles are also addressed. Furthermore, future research directions regarding SA assessment approaches are raised to deal with shortcomings of the existing state-of-the-art methods in the literature.

## I. INTRODUCTION

*Situation awareness (SA)* is defined as "the *perception* of the elements in the environment within a volume of time and space, the *comprehension* of their meaning, and the *projection* of their status in the near future" [1]. In the context of complex operational environment, SA is concerned with a person's knowledge of particular task-related events and phenomena. SA is critical for the reliable operation of almost all systems and domains [2-4]. In the aviation industry, SA is one of the key elements in pilot training for flying, controlling, and maintaining an aircraft. While most airports are intentionally built in flat terrain areas, a vast number of airports are still actually close to significantly higher terrains. Conducting instrument landing approaches into these airports imposes a high *workload*; therefore, increasing the risk of controlled-flight-into-terrain (CFIT) accidents [5]. Statistics from Boeing reveal that 50% of the total fatal accidents of worldwide commercial jet fleet from 1999 through 2008 occurred during flight descent to landing [6]. While pilot training programs have proven effective for the acquisition of the necessary technical knowledge and flying skills, non-technical skills of pilots, which include decision-making, crew cooperation, and general systems management, are also important [7]. In this aspect, non-technical competencies related to SA and threat management play a vital role in the prevention of accidents/incidents, such as CFIT and runway incursions [1]. Therefore, numerous techniques have been proposed for monitoring and assessment of SA in pilot training and air traffic control, in order to improve pilot competency and increase flight safety.

In the aviation domain, crew resource management training is now mandatory for increasing the safety of aviation operations. However, statistics indicate that 70% of air accidents/incidents worldwide are the result of flight crew actions [8]. One of the factors is that pilots take a high risk while manoeuvring at low altitudes and, indeed, 8 out of 10 general aviation accidents/incidents are caused by pilot actions [9]. Investigations show that many accidents/incidents are due to *loss of SA*. Loss of SA [1, 10] happens as the pilot's *mental model* starts to deviate from reality. The pilot does not consciously realise that a critical event has taken place. When subsequent events occur, the pilot structures the relevant events into his/her current mental model of the situation. The pilot continues to absorb information from the environment and restructure his/her mental model until the occurrence of an event that triggers a highly disconcerting awareness that the pilot's mental model of the current environment is actually false. When the pilot realises that the model is wrong, there is a collapse of this erroneous model followed by a frantic re-assessment and rebuilding of the model, if he/she is still alive and there is enough time and enough control bandwidth.

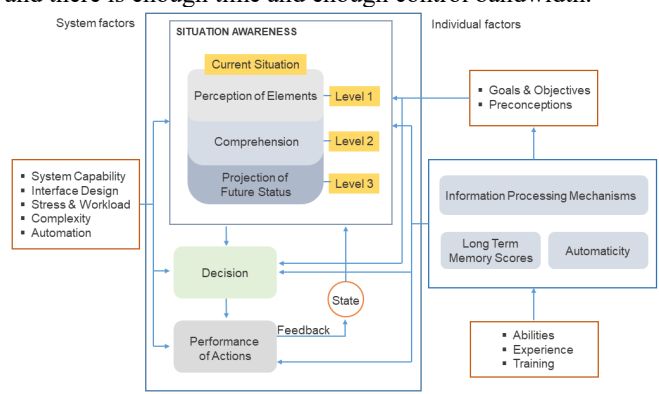

Fig. 1. The three-level model of SA, adapted from [1].

Loss of SA is one of the main causes of CFIT-related accidents. Indeed, CFIT-related accidents have caused more than 35,000 fatalities from the emergence of civil aviation in the 1920s to the turn of the 21st century [11]. From the above account, it is evident that studies on developing methods for

T. Nguyen was with the Institute for Intelligent Systems Research and Innovation, Deakin University, Waurn Ponds, Vic 3216, Australia. He is currently with the School of Information Technology, Deakin University, Burwood, Vic 3125, Australia (e-mail: thanh.nguyen@deakin.edu.au).
C. P. Lim, N. D. Nguyen and S. Nahavandi are with Institute for Intelligent Systems Research and Innovation, Deakin University, Waurn Ponds, Vic 3216, Australia.
L. Gordon-Brown is with Royal Australian Air Force Base, East Sale, Vic 3852, Australia.

assessing pilots' SA are important, as they bring valuable benefits to the aviation industry in terms of pilot training and flight safety.

According to the well-accepted SA model of Endsley [1], there are three levels of SA as illustrated in Fig. 1, i.e. perception (*level 1*), integration and comprehension (*level 2*), and prediction (*level 3*). Level 1 SA encompasses the concept of critical information observation. Level 2 SA involves the integration and interpretation of critical information. Level 3 SA is concerned with the awareness to predict possible events in the environment.

Previous reviews [12-14] have specified six main categories of SA measurement and assessment methods in the literature: freeze-probe recall techniques, real-time probe techniques, post-trial self-rating techniques, observer rating techniques, process indices, and performance measures. Based on this categorisation, this review paper provides an updated overview of recent SA assessment and measurement approaches that have been used particularly in aviation environments, including cockpits, air traffic control, and unmanned air vehicles (UAVs). The paper touches upon a wide range of critical aspects in the field, i.e. the consolidation of key SA measurement approaches, theoretical SA models and survey analyses. In the next section, six main SA categories as well as their advantages and disadvantages are explained. Different theoretical models of SA comprising individual, team and systems [15] are addressed in Section III. A survey on recent SA measurement methods is given in Section IV, which is followed by discussions and future research directions in Section V.

## II. SA Measurement Categories

### A. Freeze-probe techniques

To allow SA measurement using freeze-probe techniques during pilot training, the task performed by a subject (pilot) in a flight simulator is randomly frozen. All displays are blanked, and a set of SA queries is put forward. The subject is required to answer the queries based on his/her knowledge and understanding of the situations at the current environment ("frozen points"). The subject's responses are recorded and compared against the current status of the actual environment, in order to yield a SA score. The situation awareness global assessment technique (SAGAT) [16], SA control room inventory (SACRI) [17] – an extension of SAGAT, and SA of en-route air traffic controllers in the context of automation (SALSA) [18] are widely used freeze-probe techniques.

The freeze-probe techniques have the advantage of being a direct and objective SA measurement that eliminates the hassles involved in SA data post-trial collection, and eradicates issues related to subjective data, which are the key problems of self-rating techniques. These techniques, however, have several disadvantages that relate to the high level of intrusion on task performance (i.e. imposed by task freezes) and difficulties to implement such methods during real-world activities, especially in the circumstance of multiple subjects across multiple locations. There are also concerns with regards to their validity, e.g. memory may be assessed instead of SA, the recall of less distinct elements may not imply poorer SA, or the subject may not be aware of elements that he/she is prompted to recall. Although several improvements have been proposed to make freeze-probe techniques more applicable, it is still challenging to incorporate "freezes" into real-world exercises. This is the problem that has not yet been addressed satisfactorily in the current literature.

### B. Real-time probe techniques

A set of SA queries are administered online during task execution, but without freezing the task. Specifically, the subject matter experts (SMEs) prepare queries either before or during task execution, and administer them at the relevant points while the participant is performing the task. The answers and response time are recorded to measure the participant's SA score. The situation present assessment method (SPAM) [19] and SA for SHAPE (Solutions for Human-Automation Partnerships in European Air Traffic Management) on-Line (SASHA_L), which was introduced by Eurocontrol™ based on SPAM [20] are typical real-time probe techniques used to assess SA of air traffic controllers.

Real-time probe techniques can be applied 'in-the-field' and reduce the level of intrusion imposed by task freezes in the freeze-probe techniques. However, the extent to which the intrusion is diminished is questionable because the SA queries are still conducted online during task execution, which signifies a level of intrusion upon the primary task. The SA queries may direct participants to relevant SA information, leading to biased results [21]. Furthermore, it is difficult to apply these techniques in dynamic and unpredictable environments because SA queries must be generated in real-time and that potentially imposes a great burden upon the SMEs. Evaluating team or shared SA is also problematic using such an approach.

### C. Post-trial self-rating techniques

Each subject provides a subjective measure of his/her own SA based on a rating scale after task execution. There is a rich literature on subjective rating methods including the situation awareness rating technique (SART) [22], situation awareness rating scales technique (SARS) [23], Cranfield situation awareness scale (C-SAS) [24], crew awareness rating scale (CARS) [25, 26], mission awareness rating scale (MARS) [27], and quantitative analysis of situational awareness (QUASA) [28].

Self-rating techniques are quick and easy to use, and carried out post-trial so they are non-intrusive to task execution. Expensive simulators, SMEs and lengthy training processes are not required because of the simplistic nature of self-rating techniques, which decreases the associated implementation cost significantly. More importantly, self-ratings can be applied to assess team SA because each team member can self-rate his/her own SA performance [29]. These techniques, however, have various problems related to the collection of post-trial SA data. SA ratings may correspond to performance in a selective manner [16], i.e. subjects performing well in a trial normally rate their SA as good; while subjects are likely to forget the periods they have poor SA, and more readily recall the periods when they have good SA. Moreover, post-trial questionnaires can only measure SA of participants at the end of the task because humans are normally poor in recalling details of past mental events. Self-ratings are sensitive, e.g. subjects often rate poor SA inaccurately as they may not know that they suffer from poor SA from the beginning.

TABLE I
A SUMMARY OF ADVANTAGES AND DISADVANTAGES OF METHODS IN SIX SA CATEGORIES

| Technique categories | Typical methods | Advantages | Disadvantages |
|---|---|---|---|
| Freeze probe | SAGAT [1, 31]<br>SALSA [18]<br>SACRI [16, 17] | * Direct, objective nature, which eradicates the problems related to collecting SA data post trial. | * Intrusive method and require a simulation of the task<br>* Difficult to implement in real world activities where there may be multiple agents.<br>* Validity is questionable, as whether SA or memory is being assessed. |
| Real-time probe | SPAM [19]<br>SASHA [20] | * Perform online without freezing the task.<br>* Decrease intrusive level compared to freeze probe methods.<br>* Able to be employed 'in-the-field' activities. | * Intrusive method.<br>* Queries may direct subjects to relevant SA information, leading to biased results.<br>* Generation of probes in real-time possibly places a great burden upon SMEs, and is difficult in dynamic environments.<br>* Measurement of team or shared SA is difficult to conduct as numerous SMEs are needed. |
| Post-trial self-rating | SART [22]<br>SARS [23]<br>CARS [25]<br>MARS [27]<br>QUASA [28] | * Non-intrusive nature as they are conducted post-trial.<br>* Ease of implementation (easy, quick, require little training and low cost)<br>* No expensive simulators, SMEs or prolonged training process is required.<br>* Can be acquired from different team members and so offer a potential avenue into the assessment of team SA. | * Subjective rating<br>* Problems related to collecting SA data post-trial, e.g. correlation of SA with performance, poor recall.<br>* Sensitivity, i.e. subjects may not be able to precisely rate their poor SA as they may not realize that they have inadequate SA from the beginning. |
| Observer-rating | SABARS [27] | * Can be applied to measuring SA 'in-the-field' (real world activities) due to their non-intrusive nature. | * Concerns regarding their validity, e.g. observers may not be able to rate operators' SA; good performance may not associate with good SA.<br>* May be subject to bias because participants change their behaviors since knowing that they are being observed.<br>* Multiple SMEs are required. |
| Performance measures | Operation score [32] | * Simple to achieve and are non-intrusive because they are administered over the natural flow of the task.<br>* Can be used as a back-up SA measure for other methods. | * Problems regarding the relationship between SA and performance, e.g. an expert subject may attain satisfactory performance even when his/her SA is poor.<br>* Assume that efficient performance corresponds to efficient SA and vice versa. |
| Process indices | Eye tracker [33] | * Non-intrusive method.<br>* Can be used to determine which situational elements the subject(s) fixate upon while performing the task. | * Indirect nature<br>* Difficult to implement outside of lab settings<br>* Affected by temperamental nature of the equipment<br>* Problematic with the 'look but do not see' phenomenon by which subjects may fixate upon an environmental element but do not accurately perceive it. |

*D. Observer-rating techniques*

The SMEs provide a rating by observing each subject during task execution. The SA ratings are obtained based on observable SA associated with the subjects' behaviours in performing a task. The situation awareness behavioural rating scale (SABARS) is a typical observer rating method, which uses a five-point rating scale [27]. The main advantage of observer-rating techniques is their non-intrusive nature as they have no impact on the task being executed. These methods can also be applied to real-world activities. However, it is doubtful that observers can accurately rate the internal process of SA. A superior performance may not really equate to good SA. Some observable behaviours may suggest implication of SA, but the actual internal SA level cannot be precisely assessed by observation alone. These methods may be subject to bias because participants may change their behaviours since knowing that they are being observed. Moreover, the fact that observer-rating techniques require frequent access to multiple SMEs over a long duration is problematic, if not impossible.

*E. Performance measures*

The achievement of a subject in certain events during task execution is analysed and rated. Depending upon the task, several characteristics of performance are recorded to establish an indirect measure of SA. Performance measures are non-intrusive as they are produced through the natural flow of the task. As these methods are simple, they are normally used as a back-up SA measure of other techniques. Performance measures have a drawback, i.e. the assumption that efficient performance corresponds to efficient SA, and vice versa. An experienced subject may obtain a satisfactory performance even when his/her SA is inadequate. In contrast, due to inexperience, a novice subject may acquire superior SA levels but still attain inferior performance.

*F. Process indices*

Process indices comprise procedures that record, analyze and rate the processes that the subject follows to establish SA during the task performance. Measuring the subject's eye movements during task execution is one of the examples of process indices. Eye-tracking devices can be employed to determine which situational elements the subject has fixated upon, and evaluate how the subject's attention is allocated. Process indices have disadvantages associated with the temperamental characteristics of the equipment in their operation, and the lengthy data analysis that requires a high workload of the analyst. The use of an eye-tracking device outside of laboratory settings is not convenient. Moreover, process indices have the indirect nature, i.e. the 'look but do not see' phenomenon by which the subject may fixate upon a certain environmental element but does not accurately perceive it [30].

Table I summarizes the strengths and weaknesses of the techniques in six SA categories. Typical methods of each category are also presented along with the relevant papers.

III. TYPES OF THEORETICAL SA MODELS

The concept of SA is projected differently through different world views and their theoretical basis. Early SA models are projected in a world view that is very different from today's emerging world views. In the early days, SA under cognitive psychology perspective was dominant, and considered as the first cognitive revolution whereby cognitive systems engineering and *systems thinking* were sparsely addressed. The focus on systems thinking, however, has grown considerably today, and the SA standard concept has been shifted to the so-called second cognitive revolution [34]. The development of theoretical models of SA from the perspectives of individuals, teams, and systems levels is explained in the following sub-sections.

*A. Individual SA*

Individualistic SA models mainly focus on how individual operators acquire SA cognitively during task execution. These models rely acutely on psychology to understand the processes behind the awareness developed in the minds of individuals [35, 36]. It is difficult to explain the formative aspects of SA, and to prove the argument that knowledge in the head (or expectancy) can enable actors to create a rich awareness of their situation from very limited external stimuli. These fundamental limitations to individualistic/cognitive methods of SA can be crystalized into a set of tacit assumptions about SA itself; namely (1) it is a cognitive phenomenon residing in the heads of human operators; (2) there is a ground-truth available to be known; (3) good SA can be derived from reference to expert or normative performance standards. This is not to suggest models of SA referencing these tenets are not effective in certain practical situations. It is simply important to acknowledge that in applying individualistic models of SA, certain assumptions are being made.

*B. Team and shared SA*

Complex environments often involve multiple stakeholders; therefore, individual SA may no longer be adequate. Although the ability to evaluate SA of individual team members plays a significant role in assessing team SA, it is more important to measure the overall or shared team SA during task execution. The focus, therefore, is shifted from individual SA to team or system-level SA. Salas et al. [37] defines a team as comprising two or more participants, who possess multiple information resources and operate towards certain shared goals. Team or shared SA characterizes the coordinated awareness that the team establishes. Erlandsson et al. [38] introduced a situational adapting system to assessing team SA for fighter pilots based on information fusion. To achieve team SA, individual pilots need to develop and retain their own SA while performing the task, share their SA and notice relevant activities of other members in the team. The situational adapting system aids the pilots to cope with three different aspects: flight safety, combat survival and task completion. Collaborations between fighter pilots in a team enhance the probability of both mission success and survival.

Gorman et al. [39] introduced a theoretical framework for team SA assessment using a process-based measure namely *coordinated awareness of situations by teams (CAST)*. This measure is task based, focusing on interactions, and can be taken in vivo while tasks are performed, therefore no interruption interferes with the tasks. CAST can observe team cognition directly without the need to draw inferences about it based on accumulated individual measures. Unlike *knowledge-based measures* relying on query methods that focus on outcome of assessment by probing the memory retrieval processes of operators, CAST measure is process oriented where its output reflects the process of SA assessment [40]. This difference has an important implication when applying to decentralized command and control environment where query measures have difficulties to diagnose the processes underlying poor team SA. By capturing patterns of coordinated interactions, CAST approach alternatively is able to assess team SA by providing diagnostic information regarding the adaptability of team member connections.

Likewise, Shu and Furuta [41] developed a team SA model based on mutual beliefs taking into account characteristics of cooperative activity. This model characterizes cooperative team process more appropriately than conventional team SA definitions, which are the intersection of individual SA. With conventional team SA notion, each team member is required to have common understanding of the context, which is not realistic in cooperative team process and every team member may share the common but incorrect SA. By paralleling individual SA and mutual beliefs, the Shu and Furuta's model allows team members to complement each other, enables them to find the error immediately despite their incomplete knowledge.

Recently, Chiappe et al. [42] argued that Endsley's SA theory [43] is not able to distinguish between *weak and strong shared SA*. The latter involves mutual knowledge that improves communication and enhances team cohesion, not just overlapping representations as of weak shared SA. They proposed a situated approach to shared SA where SA comprises knowledge of actions that can be used to retrieve information embodied in the environment, instead of storing the information directly in working memory of team members. Operators use minimal internal representation by shifting information to their environment whenever possible and relying instead on regular interactions with the environment to maintain their awareness of a situation [44]. Common picture is generated by operators functioning as joint cognitive systems, which do not require extensive working memory resources and detailed world-modelling as in Endsley's theory [43]. This approach eliminates limitations with regard to working memory capacity of team members. It however has a drawback when dealing with dynamic environment where creating a common picture of a situation is greatly challenging.

*C. System and distributed SA*

Initially addressed in [45], system SA has been expanded from the increasing development of systems thinking as it presents distinctive challenges in SA assessment. These challenges can be explained starting from the *socio-technical systems (STS)* concept. STS delineates the integration of humans with the interacting technical elements to support organizational activities. Teams and team working play an important role in STS, but STS does not merely consist of teams. Indeed, STS interactively combines humans and systems, and behaves in complex, non-deterministic and often non-linear and non-additive environments. This complexity is noticeable in fields that require high technologies and critical safety standards such as aviation, aerospace, chemical, healthcare, defence and nuclear sectors. STS is actually stimulated by the challenge to these safety-critical systems [46], where SA should be distributed throughout team members, and more importantly through the artefacts employed by the participating teams. As a result, these systems emphasize on the interactions between team members and artefacts, rather than individual team member cognition.

*The Event Analysis of Systemic Teamwork* [47] is a framework of ergonomics methods that can be used to analyse performance of STS. Three network-based models including task network, social network and *SA network* constitute a "network of networks" approach to analysing systems' performance [48, 49]. Task network describes main goals and tasks performed by the system whilst social network explains the organization of the system as well as communication nature between humans and objects. SA network alternatively exemplifies the information and knowledge sharing protocol within the system, leading to more informed decisions [50].

Fig. 2 illustrates different theoretical SA models corresponding to different SA levels, where no ubiquitously superior model could be claimed. Each model suits a particular problem depending on its fundamental nature [36]. Problems addressed can be on a range from the stable, normative, individual personnel focus to the STS level in which SA is neither normative nor stable. A flexible SA method is needed to deliver the required insights with moderate analytical effort.

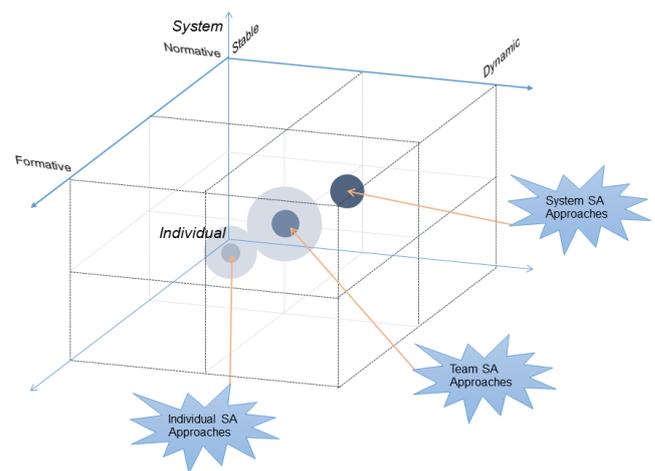

Fig. 2. Different theoretical SA models that address problems with different SA levels [15].

Stanton et al. [51, 52] presented the *distributed SA* (DSA) theory as an alternative approach to analysing and explaining SA in systems. With DSA, the cognitive processes are presumed to be distributed amongst the agents (both human and technical) in the system. By considering the information held by the artefacts (technical agents) and humans, and the ways they interact and transact, the unit of analysis is the entire STS under investigation instead of the individuals or teams of people [53, 54]. With this regard, it is important that the right information needs to be activated and delivered to the right agent at the right time regardless of whether humans or technology owns the information. Each agent's SA is different but *compatible* with one another, and collaboratively essential for the STS to accomplish its desired goal [55]. The system's DSA and each agent's SA are dynamically maintained via the transactions that allow DSA to propagate proficiently throughout the STS. *SA transactions* refer to communications between elements of the system and interactions with the environment as well as the exchange of SA-related information between elements [56].

Recently, Chatzimichailidou et al. [57, 58] put forward special types of SA called "*risk SA*" and "*risk DSA*" that refer to the presence of threats and vulnerabilities, which may affect the system safety. Risk SA can be found in individuals whilst risk DSA is a property of the STS. Risk DSA specifies that agents can be fully aware of the threats and vulnerabilities of the parts of the system under their controls and at the same time maintains a partial overview of those that could affect the entire system. This capability of the system is named *risk SA provision (RiskSOAP)* whose numerical expression can be used to quantify the positive correlation between DSA and the safety of a complex STS [58].

## IV. SURVEY OF RECENT SA APPROACHES IN AVIATION

In aviation, SA studies mainly focus on SA of pilots and air-traffic controllers. There is also an emerging research trend in SA of unmanned air vehicles. Before addressing these studies, we describe recent notable quantitative SA methods in sub-section A. Summary of advantages and disadvantages of these quantitative methods is presented in Table II. Sub-section B overviews qualitative SA methods, which are divided into three parts, corresponding to SA for pilots, air traffic controllers and unmanned air vehicles, respectively.

### A. Quantitative SA methods

Wilson [59] studied the use of *psychophysiological measures* in SA assessment. Psychophysiology explores the relationship between an operator's cognitive activity and associated changes in physiology. This kind of measures has few important properties such as unobtrusive and continuous nature that make it attractive in SA studies. French et al. [60] applied psychophysiological measures based on continuous EEG data to assess three levels of SA. Results obtained through an experiment on twelve male participants showed an acceptable accuracy of this measure in discriminating three SA levels. Gluck et al. [61] used *concurrent verbal protocol* along with retrospective verbal protocol and eye movement data to validate the control focus and performance model applied to cognitive modelling for pilots of remotely piloted aircraft. The subjects' thinking aloud talks are recorded while they operate the aircraft and the verbalizations are transcribed and coded by a coding system that comprises 22 different codes. This measure has also been used in other domains such as on-road driving [62, 63], manual-handling task studies [64], or recommendation agent trust (distrust) building process [65].

Hooey et al. [66] characterized the environment of multiple subjects by *situation elements* (SEs), which are processed through a series of sub-models such as visual attention, perception, and memory. The SA is measured by a ratio between the number of SEs that is perceived and comprehended (actual SA) and the number of SEs needed to accomplish the task (optimal SA). For every task $i$, at time $t$, actual SA is computed as the weighted sum of $m$ required SEs and $n$ desired SEs with the corresponding perception levels $p_r$ and $p_d$. They have a score of 0, 0.5 or 1, if the SE is undetected, detected, or comprehended, respectively. Specifically,

$$SA_{Actual}(t_i) = \sum_{r=1}^{m} w_r p_r + \sum_{d=1}^{n} w_d p_d \quad (1)$$

where $w_r$ and $w_d$ are the weights associated with the required SEs and desired SEs, respectively. On the other hand, optimal SA reflects the awareness when the subject comprehends all the required and desired information for task $i$, at time $t$. Therefore, the perception levels $p_r$ and $p_d$ are always equal to 1. Specifically,

$$SA_{Optimal}(t_i) = \sum_{r=1}^{m} w_r + \sum_{d=1}^{n} w_d \quad (2)$$

The SA ratio is defined as the ratio of actual SA and optimal SA. It produces a score between 0 (no SA) and 1 (maximum SA) to represent the proportion of SEs that has been aware of. Specifically,

$$SA_{Ratio}(t_i) = \frac{SA_{Actual}(t_i)}{SA_{Optimal}(t_i)} \quad (3)$$

This SA model has been validated based on a high-fidelity simulation involving a crew of two pilots performing a landing approach into an airport. The experiments demonstrate that the model is sensitive to differences of display designs and pilot responsibilities. This study constitutes preliminary results towards the development of a model-based tool to predict multi-operator SA as a function of different procedures and display designs.

On the other hand, Liu et al. [67] proposed a quantitative *attention allocation model*. The model examines the effect of information importance on SA, cognitive status of SEs and the Bayesian conditional probability theory. The sensitivity coefficient $e_i$ is used to specify the influence degree of situation element $SE_i$ on SA. This coefficient also indicates the importance $u_i$ of each $SE_i$ for the current task $j$; therefore $e_i = u_i$. The SA is then expressed through the cognitive status of SEs, as follows:

$$SA(t_j) = \sum_{i=1}^{n} e_i P_i = \sum_{i=1}^{n} u_i P_i \quad (4)$$

The attention resource allocated to $SE_i$ is defined as $A_i = \beta_i V_i Sa_i E_i^{-1}$, where $\beta_i$ represents the occurring frequency of $SE_i$, $V_i$ characterizes the information priority, $Sa_i$ signifies the salient element, and $E_i$ specifies the amount of effort to obtain the information. The attention allocation proportion of $SE_i$ is described by $f_i = A_i / \sum_{i=1}^{n} A_i$. If the activity that the subject pays attention to $SE_i$ is described as event $a_i$ at time $t$, the occurrence probability of $a_i$ is identical to the attention allocation proportion, i.e.,

$$p(a_i) = f_i \quad (5)$$

A high-level of SA is dependent on the lower levels of SA; therefore if event $a_i$ has occurred, event $b_i$ that $SE_i$ would not be comprehended may occur with probability $k_i = p(b_i/a_i)$, and event $c_i$ that $SE_i$ would be comprehended may occur with probability of $p(c_i/a_i) = 1 - k_i$. Therefore, the expectancy of the cognitive level of $SE_i$ can be computed by the Bayesian theory:

$$\bar{p}_i = (1 - p(a_i)) * 0 + p(a_i b_i) * 0.5 + p(a_i c_i) * 1 \quad (6)$$

At time $t$, the SA can be expressed through the attention allocation as follows:

$$SA(t_j) = \sum_{i=1}^{n} u_i \bar{p}_i = \frac{\sum_{i=1}^{n}(1 - 0.5 k_i) u_i A_i}{\sum_{i=1}^{n} A_i} \quad (7)$$

Experiments involving 20 pilots show that the SA predicted by this model is greatly correlated with the operation performance as well as measurement indices such as the correct rates of SAGAT, 3-D SART, and physiological features, e.g. pupil diameter, blink frequency, or eyelid opening. It thus demonstrates that the SA model is useful for predicting the changing trend of SA during task performance.

The attention allocation model has been extended and optimized in Liu et al. [68], by considering a *cognitive process analysis* to understand the internal process of SA based on the basic theories of adaptive control of thought-rational (ACT-R). The relationship between the ACT-R cognition theory and a pilot's three levels of SA is demonstrated in Fig. 3.

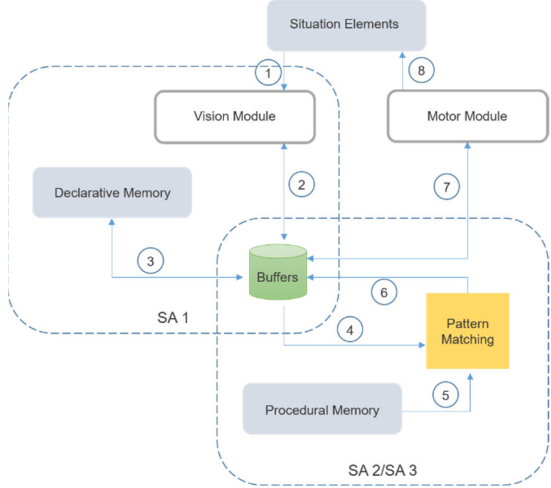

Fig. 3. Three levels of SA based on ACT-R: (1) determine what and where to see, (2) obtain the visual information of situation elements, (3) retrieve the chunk, (4) match the IF part, (5) select the rule, (6) execute the THEN part, (7) prepare for movement, and (8) carry out movement.

The *vision module* selects and registers certain SEs into the short-term sensory store after they are filtered by the selective attention. Then the *buffers* obtain situational information from the vision module, and access the *declarative memory* module to retrieve the corresponding knowledge. SE is perceived at the first level ($SA\ 1$) when its activating quantity is greater than a certain threshold. Once the "if" part of the *procedural memory*, which contains production if-then rules, matches the content of the buffer, the *pattern matching* procedure selects the corresponding rule to execute the "then" part through the *motor module*. In a dynamic system, there is a fuzzy boundary between $SA\ 2$ and $SA\ 3$ because the comprehension of the current status has direct implications on the prediction of the future, and both are equally relevant to the task. The details of this method are shown in Fig. 4, followed by mathematical formulation as below.

When the vision module determines to "see" $SE_i$ as event $a_i$ occurs, the buffers activate the corresponding chunk $i$ to $SE_i$ in the declarative memory with the level of activation $AC_i$ specified by:

$$AC_i = AC_{0i} + \sum_j W_j S_{ij} \quad (8)$$

where $AC_{0i} = 0.5 \ln t$ is base-level activation, specifying the fact that recognizing $SE_i$ (Fact $F_i$) has been presented for $t$ times; $W_j$ indicates the attention weighting of $SE_j$ due to the current Fact $F_i$; and $S_{ij}$ represents the strength of association from the current Fact $F_i$ to the related $SE_j$:

$$S_{ij} = S - \ln(fan_j) \quad (9)$$

where $fan_j$ is the number of facts associated with $SE_j$, and $S$ equals to 2 as determined in [69].

Once $AC_i$ is greater than threshold $\tau$, the SE is perceived at level 1 (SA 1), and is considered as event $b_i$:

$$p(b_i|a_i) = \left(1 + e^{-(AC_i-\tau)/s}\right)^{-1} \quad (10)$$

where $\tau = 1.0$ and $s$ is typically set to 0.4 for controlling the noise in the activation levels [69]. According to the ACT-R theory, multiple production rules are possibly matched at any time, but only the rule with the greatest utility $U_i$ is selected for execution. The current SE is fully comprehended either in the form of its current meaning (SA2) or the future one (SA3), which is considered as event $c_i$:

$$p(c_i|b_i a_i) = \frac{e^{U_i/\theta}}{\sum e^{U_i/\theta}} \quad (11)$$

Based on the mathematical expectancy of cognitive level $\bar{p}_i$ in (6), the current SA is computed by:

$$SA = \sum_{i=1}^{n} e_i \bar{p}_i =$$
$$= \sum_{i=1}^{n} \left(\frac{e^{U_i/\theta}}{\sum e^{U_i/\theta}} + 0.5\right) \times \frac{u_i f_i}{1 + e^{-(AC_i-\tau)/s}} \quad (12)$$

To verify this model, Liu et al. [68] used four methods to measure SA, including the freeze-probe technique (i.e. SAGAT), post-trial self-rating (i.e. 10-dimensional SA rating technique – 10D SART), performance measure (i.e. operation score) and process indices (i.e. eye movement including pupil diameter, blink frequency, and ratio of saccades). The empirical results show that the proposed SA model is useful for guiding a new design of cockpit display interfaces to reduce pilot errors [70].

*B. Qualitative SA methods*
*1) SA for pilots:*
Better aircraft designs and better flight training are important processes that help increase SA capability of pilots. Bolstad et al. [71] carried out several experiments to assess effectiveness of six training modules for developing and maintaining SA of general aviation pilots. They used the general aviation aircraft version of the SAGAT, installed on a personal computer next to the simulator to measure SA. The simulation was halted at four randomly chosen times, and the subjects were inquired to respond to SAGAT queries without looking at the displays or referring to other information. Alternatively, Sorensen et al. [72] discussed three theoretical frameworks including psychology, engineering and systems ergonomics for understanding and enhancing a pilot's SA. Although psychology and engineering provide valuable contributions to understanding SA, they both do not consider the interaction between the individuals, artefacts, and the context within which they exist. In contrast, the systems ergonomics perspective takes a holistic approach to analyse SA by considering the interactions between the individuals and artefacts as well as their environment.

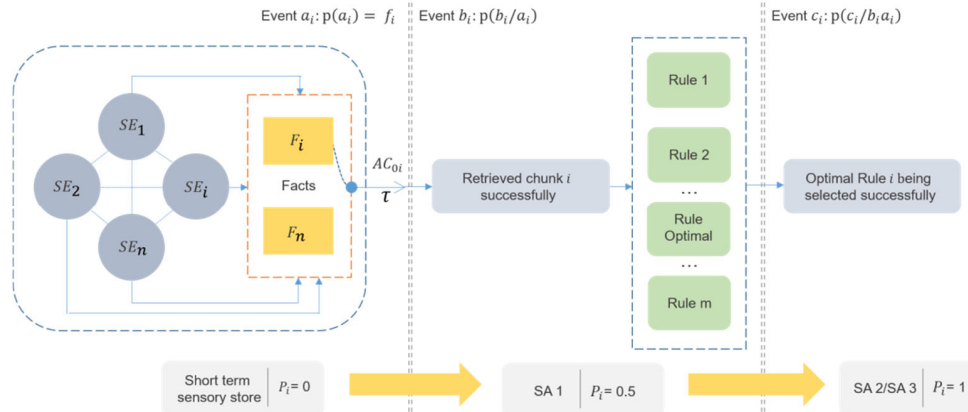

Fig. 4. SA analysis model where $SE_i$ is the $i$th situation element; $F_i$ is the fact of recognizing $SE_i$; $p(a_i)$ is the probability of paying attention to $SE_i$; $f_i$ is the attention allocation proportion; $p(b_i|a_i)$ is the probability of chunk being retrieved successfully; $p(c_i|b_ia_i)$ is the probability of optimal rule being chosen; $AC_{0i}$ is the activation level of chunk; $\tau$ is the activation threshold; and $P_i$ is the cognitive level.

TABLE II
ADVANTAGES AND DISADVANTAGES OF QUANTITATIVE SA METHODS IN AVIATION

| Quantitative SA Models | Advantages | Disadvantages |
|---|---|---|
| Psycho-physiological measures [59, 60] | * Can be EEG data, eye blinks, or heart rate.<br>* Used to study complex cognitive domains, e.g. mental workload and fatigue.<br>* Unobtrusive and continuously available.<br>* Useful in determining SA in field settings. | * Require expensive systems to obtain data.<br>* Complicated data acquisition and analysis process. |
| Concurrent verbal protocol [61-65] | * Can monitor how the pilots/subjects allocate their attention among instruments.<br>* Provide a source of high-density data. | * Require knowledge of subject matter experts.<br>* Complicated data analysis process. |
| Man–machine Integration Design and Analysis (MIDAS)-based SA model [66] | * The environment is broken down into situational elements.<br>* Allow for the prediction of SA as a function of the operator's high-level tasks. | * Only focus on the first of Endsley's three stages of SA, i.e. the perception of elements in the environment. |
| Attention allocation model [67] | * Useful for predicting the changing trend of SA during task performance.<br>* SA prediction results are correlated with the correct rates of SAGAT, 3-D SART, and physiological features. | * Limited in studying level 1 and level 2 SA only.<br>* The mental model module is complex and uncertain. |
| Adaptive Control of Thought, Rational (ACT-R) Theory-based SA model [68] | * A joint qualitative and quantitative model based on an extension of the attention allocation model [67].<br>* Considering a cognitive process analysis to understand the internal process of SA.<br>* Useful for guiding a new design of cockpit display interfaces to reduce pilot errors. | * Operators may focus only on obtaining higher correct rate rather than higher performance score, which may lead to performances that are not highly correlated with SA model. |

SA assessment using observer and self-rating methods under extremely stressful and challenging training conditions was investigated in [73]. The results show that subjective SA measures are unlikely to produce valid estimates of SA under extreme conditions. van de Merwe et al. [74] set up a flight simulator to assess SA through eye movements in a series of experiments. A malfunction (a fuel leak) was introduced to hamper SA, and the subjects (pilots) should discover the problem. The study used fixation rates and dwell times as an indicator of level 1 SA, and entropy, which is a form of randomness of scanning behavior, as an indicator of level 3 SA. Likewise, a rule-based method to assess a pilot's SA through monitoring pilot eye movement was suggested in [75]. The experiments demonstrate that eye movement of an experienced pilot is considerably different with that of novice pilots.

Wei et al. [76] studied SA in a pilot-aircraft system to improve the cockpit display interface design using a cockpit flight simulation environment. The experimental setup consisted of a virtual instrument panel, a flight visual display, and the corresponding control system as illustrated in Fig. 5. A human-in-the-loop experiment was carried out to measure SA by the SAGAT method. The SA degrees and heart rate data of the subjects were obtained, which yielded an objective evaluation of the display interface designs and provided a useful reference for aircraft designs. The experiments performed on the Boeing, Airbus and ARJ21 display interfaces reveal that the cockpit display interface design has an important impact on the SA degree of pilots.

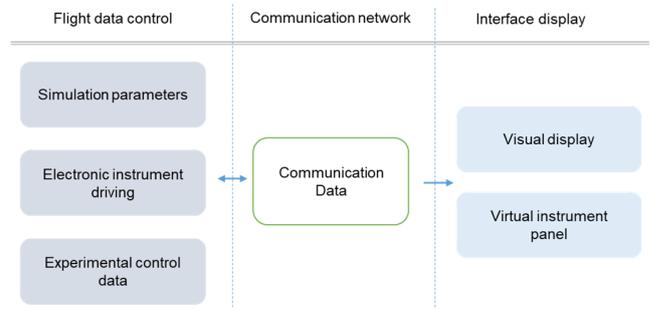

Fig. 5. Structure of cockpit flight simulation environment used in [76].

Pilots often change their levels of supervisory control between a full autopilot and other modes, and the transitions must not involve an unsafe reduction in flight performance or unacceptable change in workload or SA. Hainley et al. [77] quantified a pilot's flying performance, workload, and SA during several types of automation mode transitions, in order to develop objective measures of mode transition gracefulness. SA changes were assessed by analyzing tertiary tasks, verbal callouts of altitude, fuel, and terrain hazards. The experiments performed by thirteen subjects employed a fixed base simulation of the NASA Constellation Program Altair lunar lander. The results show that the *mental workload* increases, and SA decreases monotonically with the number of manual control loops the pilot has to close as a result of flight mode transitions. The study also highlights the loss of pilot attention to fuel status, altitude, and terrain during high workload periods due to the attentional demands of the manual control tasks.

Yu et al. [78] proposed the use of eye-tracking devices to capture pilots' visual scan patterns and measure SA performance. The research involved eighteen male military pilots, who qualified as mission-ready with flying experience varying between 310 and 2920 hours. Stepniczka et al. [79] studied the impact of stress on *social cognition* (*social SA*) with a focus on empathy, one of the key skills of social SA. The research found that stress, i.e. challenging and surprising situations during flight, led to a significant reduction in empathic accuracy and thus social SA of both pilot flying and pilot monitoring. These findings suggest changes to pilot training, procedures and system design to support crews to better manage their workload during stressful events, leading to enhanced empathic accuracy and improved crew interactions. Recently, the effects of pilot ages and flight hours on SA and workload were studied in [80]. The SPAM technique was used to measure real-time SA for 22 volunteer test pilots who were flying along with a checking pilot from a traditional UH-1H helicopter fleet. Before the post-flight debriefing, the SART and NASA task load index post-trial subjective self-rating were implemented to verify the SPAM data. The results indicate that there are correlations between SA with pilot's ages and flight hours. The findings are helpful for mission scheduling and staff management.

*2) SA for ground air traffic controllers:*

Air transport operations normally require various personnel on the ground to observe and supply weather conditions, air traffic, and other relevant information to pilots through air traffic control, automatic terminal information service, or on-board instrumentation. As an example, the ground staff may monitor weather and advise the pilots before the airplane enters a hazardous weather region. In this circumstance, the ground staff may exchange a data link message to the pilots that reports the upcoming weather or recommends a different route (e.g., a modified flight path, flight level, or destination) to circumnavigate the harsh weather region. The pilots, however, are often unable to autonomously analyse the information being relied on by the ground staff; therefore, lack the required SA to deal with the operation of the airplane. Kommuri et al. [81] proposed methods for monitoring an aircraft and providing flight tracking information to aircraft for improved SA. A typical method involves capturing a flight tracking image associated with the aircraft and communicating the image to the pilots for display on a device on-board the aircraft. The communication can also include textual information pertaining to the image, a map including a graphical representation of a region, a navigational reference point overlying the terrain or meteorological region identified by a weather monitoring system.

Vu et al. [82] studied SA and workload of air traffic controllers and pilots who participated in a human-in-the-loop simulation with various distributed air-ground traffic management concepts involving human controllers, pilots, and automation. Eight pilots began the scenario in the en-route phase of flight and were asked to circumvent convective weather region while conducting spacing and merging tasks in conjunction with a continuous descent approach into an airport. Two controllers managed the sectors through which the pilots controlled their aircrafts, with one handling a sector that comprised the top of descent, and the other supervising a sector that contained the merging point for arrival into the airport. The simulation results show that when the controllers are in charge and actively engaged, they exhibit higher workload levels than those of the pilots, and their changes in SA levels are dependent on the sector characteristics. Instead, when the pilots are in charge of separation, they have higher levels of SA, but not essentially higher workload levels. Moreover, the pilots tend to ask traffic controllers for assistance when their solutions or those from the auto-resolver mechanism fail. In these circumstances, the controllers, however, show limited awareness of the aircrafts, and normally are unable to help the pilots to resolve conflicts.

Monitoring automation leads to complacency and loss of SA, suggesting that operators become more complacent, and over-reliant on the automated tools. Mirchi et al. [83] investigated the levels of air traffic controller's trust in automation and their relationship to SA. The study was carried out over a 16-week internship involving twelve student air traffic controllers. The results indicate that the participants who score lower on the trust scale have higher SPAM probe accuracy during high traffic density scenarios while those scoring higher in trust have lower SPAM probe accuracy. One possible explanation could be that subjects with high trust scores become more complacent with automation, and this negatively affects their SA.

Chiappe et al. [84] emphasized the importance of SA to operator performance in the *next generation air transportation system*. They proposed a situated approach, i.e. *situated SA*, to understanding how individuals and teams obtain and maintain SA. The operators are suggested to rely on interaction with external tools to off-load information processing and storage, rather than relying on highly detailed internal representations and world-modeling that overload their working memory. Although the situated SA emphasizes off-loading, it also highlights that operators must not be excluded by automation tools. Unlike the distributed approach, the study argues that humans are privileged cognitive agents who must retain meaningful involvement with the props and tools characterizing their task environment. On the other hand, Blasch [85] demonstrated the use of different visualization methods to enhance SA in air-ground coordination through collecting, fusing and presenting air, ground, and space data. The developments in big data processing, e.g. machine learning, visual, and text analytics are needed. These developments provide air traffic controllers and pilots better understanding of information available in the environment, e.g. weather, airspace/airports, aircraft takeoff and landings. It is however important to note several problems with big data solutions. Firstly, there are problems relating to time dependencies, ontological discrepancies, and data completeness and accuracy. Then there are the issues to analytic treatment to get the data down to the capabilities of the human senses; and then there are the modelling and simulation issues for level 3 SA, all needing to happen in real time. Much further work is required with this SA research direction.

Visual attention is considered as a high-impact perception for air traffic controllers to accurately and efficiently perform tasks that require high cognitive workloads. Salience is one of the user interface components attracting the operators' visual attention. Yoshida et al. [86] investigated the relationships between the degree of salience gaps and the performance of air traffic controllers' tasks to develop a screen design policy based on these relationships. The simulation results show that a larger salience gap between important and unimportant airplanes leads to a shorter reaction time for responding SA queries. Better instruction timing can be obtained if the salience gaps among aircrafts are larger.

*3) SA for unmanned air vehicles:*
Operations of unmanned systems often rely heavily on human operators' interaction and control as many of these systems are tele-operated or semi-autonomous. Unmanned vehicles (UVs) are normally placed at a remote location that limits the human's understanding of the UVs and their surrounding environment. In addition, UV sensing capabilities are generally limited, or inaccurate as compared with humans' rich sensing capabilities. There are circumstances where multiple personnel are required to support a single UV or UV missions involve tasks by humans that need a demanding cognitive attention. Future UV systems will be fully autonomous and such systems with human-like reactive capabilities require UVs to possess SA.

Studies on SA so far have focused merely on the human's ability to attain and maintain SA. SA concepts involving UVs in general or UAVs in particular may be classified into two categories: the SA of the UV operators (humans) and the SA of the UVs themselves. Whilst the human aspect of SA has been examined in previous sections, the following discussion focuses on the principle how a UV may possess human-like SA capabilities, and presents a survey of recent studies related to the SA ability of UAVs.

Adams and Freedman [87, 88] introduced the concept of unmanned vehicle SA and hypothesized that a UV that possesses human-like SA capabilities can increase the mission success of future UV systems and also support the human's SA. The rich human SA literature can shape the development of the UV SA architecture mechanisms, measurement methods, as well as assessment criteria. However, a one-to-one mapping between human SA and UV SA does not exist. In fact, UV SA with the assistance of many existing artificial intelligence and perception technologies does have many superior capabilities over human SA, spanning across three levels of SA.

At the perception level, UVs can outperform humans with regard to monitoring tasks in 3D environment, i.e. dull, dirty and dangerous. Unlike humans, UVs may possess impeccable memory recall because they can store all observed information during the entire mission. With the higher information collection and working memory capacity, UVs offer higher perception capabilities than humans. Moreover, the UV sensing technology has a potential to overcome human sensory drawbacks, e.g. night vision, and auditory perception beyond the human perceptual range.

At level 2, human SA may be improved through training and experience, which can lead to a mental model of the mission. However, individuals unfamiliar with the situation have a struggle to attain the mental models of experienced persons, and this results in lower performance levels. UV SA with the comprehension transfer capabilities between unmanned entities can eliminate or reduce the novice user phenomena of human SA.

At level 3, human projection can be accomplished based on observing and understanding the situation that reflects excellent perception (SA 1) and comprehension (SA 2). Human level 3 SA is limited because prediction is often a highly demanding cognitive activity, which is affected by various aspects such as cognitive workload, mental capacity, and environmental stressors. The ability to process larger amounts of information as well as the incorporation of working-memory, long-term memory, decision making, and mental models into UV technologies can eliminate this limitation of level 3 human SA.

McAree and Chen [89, 90] proposed a method of spatial projection of traffic vehicles to provide increased *artificial SA* for an unmanned aircraft system (UAS) operating in a terminal area where heavy traffic heads to the same airfield. A highly autonomous UAS must possess the same level of SA with a manned aircraft to maintain an equivalent level of flight safety. Traffic vehicles are assumed to follow a predefined route through the terminal region, but this does not happen perfectly. The uncertainty of aircraft navigation occurring when following the nominally prescribed path is captured by utilizing a curvilinear reference frame, and is dealt with by incorporating both discrete and bounded uncertainties. This method yields significant performance benefit although it increases the computational complexity of the problem as compared with linear methods.

Human operators are responsible for the SA of the traditional UAS flight as it normally takes place under the visual line of sight (VLOS) of the human. When UAS operates beyond visual line of sight (BVLOS), the operator is no longer able to maintain unaided eye contact with the UAS, therefore this responsibility shifts between human and the increasingly autonomous vehicle. The artificial SA system of a UAS in BVLOS must be as good as SA of a human operator. Recently, McAree et al. [91] introduced a framework to probabilistically quantify the SA of a UAS, both with and without a human in the control loop. The experiments show the benefits of the proposed framework in assessing both VLOS and BVLOS systems. Eventually, the probabilistic framework allows the integration of emerging technologies in the development of artificial on-board SA systems.

Fig. 6 presents the relationship between SA of humans and UVs with the level of *autonomy*, which is a mechanism to reduce the cognitive demands imposed upon humans. It has been shown that SA can be decreased as the system *autonomous level* increases [92].

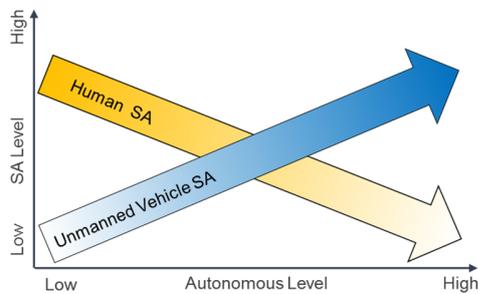

Fig. 6. The allocation of human and unmanned vehicle SA through different autonomous levels [87].

UV SA is required to correlate with the UV's level of autonomy. At the lowest autonomy level, the human has full control of the UV and the responsibility of SA is placed on the human whilst the UV has limited or no SA. However, human's SA may be improved if UV possesses some SA. As the autonomy level spans from low to high, the UV SA level increases while the human's SA level decreases. At the highest level of autonomy, the system is required to possess a high SA level to ensure safe and successful mission completion while the human may not require a high SA level.

## V. DISCUSSIONS AND FUTURE RESEARCH DIRECTIONS

Failure to monitor or observe the available information constitutes a majority of SA errors in aviation. These errors arise when the relevant data are limited, hard to discriminate or detect, or when the accessible information is perceived incorrectly, or when memory loss occurs. Enhancing SA through designing better aircraft systems and pilot training has resulted in significant impacts. In addition, an accurate SA assessment is critical as it leads to fundamental understanding of the behaviors of novice pilots in flight control. Consequently, effective guidelines and training programs can be administered for enhancing the performance of pilots and improving flight safety in the aviation industry.

This paper reviews recent SA assessment approaches including quantitative and qualitative aspects as well as theoretical models based on team and distributed or system level SA. We have discussed the advantages and disadvantages of six categories of SA assessment methods including freeze-probe techniques, real-time probe techniques, post-trial self-rating techniques, observer rating techniques, process indices, and performance measures. Despite their limitations, none of the mentioned methods has been discarded by researchers and practitioners in aviation or air traffic control. The combination of several measures when measuring SA is recommended to ensure concurrent validity. Table III summarises the relevant publications of different SA aspects.

TABLE III
SUMMARY OF RECENT SA STUDIES

| SA models | Pilots in cockpits | Air traffic controllers | UAVs |
|---|---|---|---|
| Qualitative | [31] [71]-[80] [93][94] | [18][21] [81]-[86] [93] | [87] [88] |
| Quantitative | [61][66]-[68] [70] | [59][60] | [87][91] [95] |
| Shared or team | [13][37] [38][41] [42] | [13][39] [41][42] | [87][95] [99] |
| Distributed or system | [3][15] [51][52] [54][57] | [51][52] [54][58] | [51][87] [89][90] [95][99] |

In essence, SA differs among individuals as a function of task expertise. As an example, it has been shown that novice pilots tend to be less proficient in anticipating future aircraft states [96]. This phenomenon is, in part, owing to their inflexible visual scan patterns on instrument in the cockpit. In contrast, experienced pilots are more flexible in their visual scanning capability, and their control over perception of information and interpretation of the perceived information aids them in event prediction. It has been established that individual differences in attention and memory closely affect SA [97, 98], and novice pilots differ significantly in their constructs in deriving SA. Consequently, research on *personalised SA* assessment model that considers individual differences in SA assessment is encouraged. An example of such research would combine both cognitive, e.g. electroencephalography (EEG) analysis, and physical information, e.g. eye movement, facial and gait analysis, of pilots as a useful method for SA assessment. How pilots make use of the available real-time information to sense, predict, and act in response to various adverse flight scenarios can be monitored cognitively through EEG signals and physically through eye movement in observing flight instrument and the corresponding facial and gait analysis. Additionally, some of the biological variables of importance, such as fatigue, sustained and focussed attention, and circadian rhythms that affect SA and overall decision-making would be worth investigating. The cultural effects of team-working within the cockpit, with air traffic control, and beyond are also significant. It is important to note that if SA is going to feature in engineered solutions, these aspects should be included in the solution.

Notably, we have addressed SA issues of unmanned vehicles or artificial SA and its correlation with the increasing autonomous levels of unmanned systems. Although team or shared SA methods have been investigated widely in the literature for human teaming, relatively fewer studies on interactive collaboration between humans and machines have been conducted. Indeed, *human-autonomy teaming* is an emerging trend in SA research where many relevant issues need to be investigated, e.g. shared SA, transparency, effective team communication, trust, timing, SA overconfidence and machine ethics [99]. Several research questions can be raised from this perspective. For example, can future UVs possess adjustable autonomy where the autonomous level will change based on the situations? Then, how will the UV adjustable autonomy impact the features of SA architecture? Given the advent of advanced automation and artificial intelligence, how do the SA assessment requirements change? How to address ethical issues arising from the collaboration between humans and autonomous agents? In general, it is possible to design SA architecture of unmanned systems or human-machine systems, but this must be conducted through a holistic, system-of-systems approach that integrates technologies from various fields, including artificial intelligence, autonomy, and cognitive systems. It is also worth pointing out that whatever the engineered solution looks like, and whatever levels of automation and artificial intelligence are included in this solution, people will still be liable and responsible for the behavior and consequences of the solution in operation. In addition, these people need to be able to influence the operations through informed command and informed consent, and SA is vital for both of these control modes whether the people are 'in the loop' or 'on the loop'.

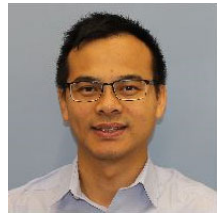
**Thanh Nguyen** received the Ph.D. degree in mathematics and statistics from Monash University, Australia, in 2013. He is currently a Senior Lecturer with School of Information Technology, Deakin University, Burwood, Vic 3125, Australia. Dr. Nguyen was a recipient of an Alfred Deakin Post-Doctoral Research Fellowship in 2016.

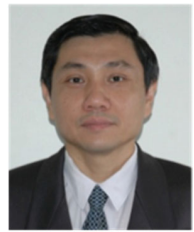
**Chee Peng Lim** is currently a Professor at the Institute for Intelligent Systems Research and Innovation, Deakin University, Australia. He has more than twenty years' experience in trusted autonomy and computational intelligence research, especially in theoretical development and practical application of data-based learning models.

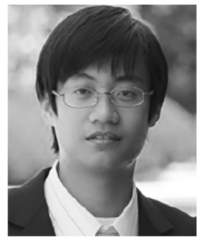
**Ngoc Duy Nguyen** received the M.S. degree in computer engineering from Sungkyunkwan University, Seoul, South Korea, in 2011. He is currently pursuing the PhD degree in Information Technology with the Institute for Intelligent Systems Research and Innovation, Deakin University, Australia.

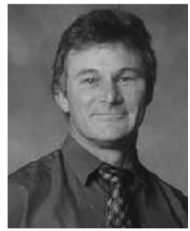
**Lee Gordon-Brown** received the B. Eng. (Mech) from Swinburne University of Technology, Australia, and the PhD degree from Monash University, Australia. Dr. Gordon-Brown is currently working for Royal Australian Air Force Base, East Sale, Victoria, Australia.

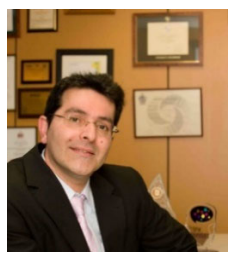
**Saeid Nahavandi** received a Ph.D. from Durham University, U.K. in 1991. He is an Alfred Deakin Professor, Pro Vice-Chancellor, Chair of Engineering, and the Director for the Institute for Intelligent Systems Research and Innovation at Deakin University.

His research interests include modelling of complex systems, robotics and haptics. He has published over 600 papers in various international journals and conferences. He is a Fellow of Engineers Australia (FIEAust), the Institution of Engineering and Technology (FIET) and Senior Member of IEEE (SMIEEE).